\documentclass[12pt]{article}
\usepackage[koi8-r]{inputenc}
\usepackage[english]{babel}
\usepackage{psfig}

\begin{document}
\begin{center}
{\Large \bf
Observations of Lick standard stars using the SCORPIO 
multi-slit unit at the SAO 6-m Telescope}\\
\bigskip

{\large Sharina M.E.$^1$$^*$, Afanasiev V.L.$^1$, Puzia T.H.$^2$}

\bigskip

{\em $^1$ Special Astrophysical observatory Russian Academy of Sciences,
N.Arkhyz, KChR, 369167, Russia

 $^2$ Space Science Telescope Institute, 3700 Sun Martin Drive, Baltimore, MD21218, USA}
\end{center}

\begin{abstract}
We present Lick line-index measurements of standard stars from the list of Worthey (1994).
The spectra were taken with the multi-slit unit of the SCORPIO spectrograph at the 6-m Special
Astrophysical observatory telescope.
We describe in detail our method of analysis and explain the importance
of using the Lick index system for studying extragalactic globular clusters.
Our results show that the calibration of our instrumental system to the standard Lick system
can be performed with high confidence.

\end{abstract}
-----------------------------  \\
$^*$e-mail: sme@sao.ru
\newpage

\begin{center}
{\large \bf Introduction}
\end{center}

\bigskip

\noindent
Studying the line-strength indices plays an important role in  analysing the
ages, metallicities, and [$\alpha$/Fe] ratios of unresolved distant
stellar populations. The behavior of indices in composite stellar systems
has supplied fundamental clues to the understanding the star formation
histories of galaxies (e.g. Faber 1973, Burstein et al. 1984).

Globular clusters (GCs) are simple stellar systems composed of stars  of
one age and chemical composition and offer a unique tool to access  the
star formation histories of their own galaxies. GCs are gravitationally
bound objects, whose lifetimes may exceed the Hubble time. They are the
most luminous simple stellar populations known to exist. If the
correlations between key spectral indices, metallicities and ages are
common for GCs in all types of galaxies, we can determine the age for the
given particular GC by comparison with evolutionary population synthesis
models (e.g. Worthey 1994, Vazdekis 1999, Bruzual \& Charlot 2003, Thomas
et al. 2003).

We observe GCs 
 in nearby galaxies using the multi-slit unit of the
SCORPIO spectrograph, mounted at the prime focus of the 6m telescope. Our
ultimate objective is to study the star formation  and chemical enrichment
histories of globular cluster systems and their host galaxies. However,
first, we will check the correspondence of our instrumental system of
line-strength indices to the Lick standard system. In the following we
study the influence of random and systematical errors arising in the
process of observation and data reduction on  the Lick index measurements.

\vspace{1cm}

\begin{center}
{\large \bf Spectrophotometric system of Lick indices.}
\end{center}

\bigskip

\noindent
\begin{center}
{\bf \it  Definitions}
\end{center}
Understanding the physical origin of prominent absorption features in the
integrated light of simple and composite stellar systems (i.e. star
clusters and galaxies) is based on an empirical study of variations of these
features in Galactic stars with effective
temperature, mass, and chemical composition. For this, Worthey et al.
(1994) measured indices for a sample of 460 Galactic stars. Spectra in a
range of 4000--6000 \AA\ were obtained in the Lick observatory in
1972-1984 using the same instrumentation. Spectral resolution of the Lick
system is 8--10 \AA\ (Worthey \& Ottaviani 1997). Information
about the measured indices for stars with various  $log g$, $T_{\rm
eff}$ and $[Fe/H]$ is used to construct the fitting functions, which are the main
ingredients for computing theoretical models of simple stellar populations.

Despite the existence of different approaches to measuring the
line-strength indices, most of the authors adhere
to the classical definition of the indices as an analog of the
equivalent width (e.g. Faber 1973, Puzia et al. 2002):
$$ W_{\lambda} = \int_{ \lambda_{min}}^{\lambda_{max}}
\left(1 - \frac{F_l( \lambda)}{F_c( \lambda)}\right) d \lambda, \eqno(1) $$

where $F_l(\lambda)$ is the observed spectrum and $F_c(\lambda)$ is the
local continuum, which is usually determined via interpolating 
$F_l(\lambda)$ between two neighbouring spectral ranges on the red and blue sides of the line.

The so-called theoretical way of the index calculation 
$$ I_t(\lambda) = \left(1 - \frac {\int F_l(\lambda) d\lambda}{ {\int F_c(\lambda)
d\lambda}}\right) \cdot \Delta \lambda, \eqno(2) $$
does not differ strongly from the ''observational'' definition (1) for
spectra with a high $S/N$ ratio. There are no systematic differences between
the two definitions. The scatter is $<0.1$\%. $I_t(\lambda)$ is the most
common definition and yields more plausible results for spectra with low $S/N$
($\le10$ per \AA). The observational definition is commonly used in the literature.

In order to avoid subjectivity in drawing the local continuum, the
line-strength indices are completely characterized by three wavelength
ranges: one central one, which includes the spectroscopic feature, and two
neighbouring ones located towards the red and blue of the central region
\\ (http://astro.wsu.edu/worthey/html/index.table.html). Geisler (1984)
and Rich (1988) pointed out that a pseudo-continuum is measured instead of
true continuum at low spectral resolution. Molecular line-strength indices
are usually measured in magnitudes:
$$I_m(mag)=-2.5\cdot\log_{10}\left(1-\frac{W_{\lambda}({\rm \AA})}{\Delta
\lambda}\right), $$ where $\Delta \lambda$ is the width of the central
wavelength range. Molecular indices are commonly used to measure broad molecular
features in spectra. The continuum regions for them are located far from the
central region.

Atomic indices are measured in \AA ngstroms:
$$ I_a({\rm \AA}) = \Delta \lambda (1-10^{-0.4\; I_m}).$$ In that case,
the lines and the central wavelength regions have narrower width.
The continuum regions are closer to the central bandpasses.

\bigskip

\begin{center}
{\bf \it Error estimation in the measurement of the Lick indices}
\end{center}

Obviously, the suitability of line-strength indices to investigate the
above mentioned items relies on the proper determination of the associated
index errors.

There are the following sources of systematic errors
in the line-index measurements:
flux calibration uncertainties,
corrections from spectral resolution and velocity dispersion,
sky subtraction uncertainties, scatter light effects,
wavelength calibration and radial velocity errors, seeing and focus
length variations, deviations from linearity, and contributions
from nebular emission lines. The statistical errors arise
from the random Poissonian noise.

\begin{center}
{\large \bf Observations and data reduction}
\end{center}

\bigskip

\noindent
\begin{center}
{\bf \it Instrumental characteristics}
\end{center}
The observations were performed with the multi-slit unit of the SCORPIO
spectrograph  (Afanasiev \& Moiseev 2005). In this mode SCORPIO has 16
movable slits (1.2" x 18") in the field of 2.9 x 5.9 arcminutes in the
focal plane of the telescope. We use the CCD detector EEV42-40 with 2048 x
2048 pixel elements and the scale $\sim\!0.18$ arcminutes per pixel. The
holographic grism VPHG1200g (1200 lines/mm) gives a spectral resolution
$\sim\!5$ \AA. The exact covered spectral region depends on the distance
from the center of the field. Most of the spectra have good wavelength
coverage between 4000--6000 \AA.

The original, non-reduced spectra obtained with SCORPIO in multi-slit mode are
shown in Fig.~1.
\begin{figure*}[!th]
\psfig{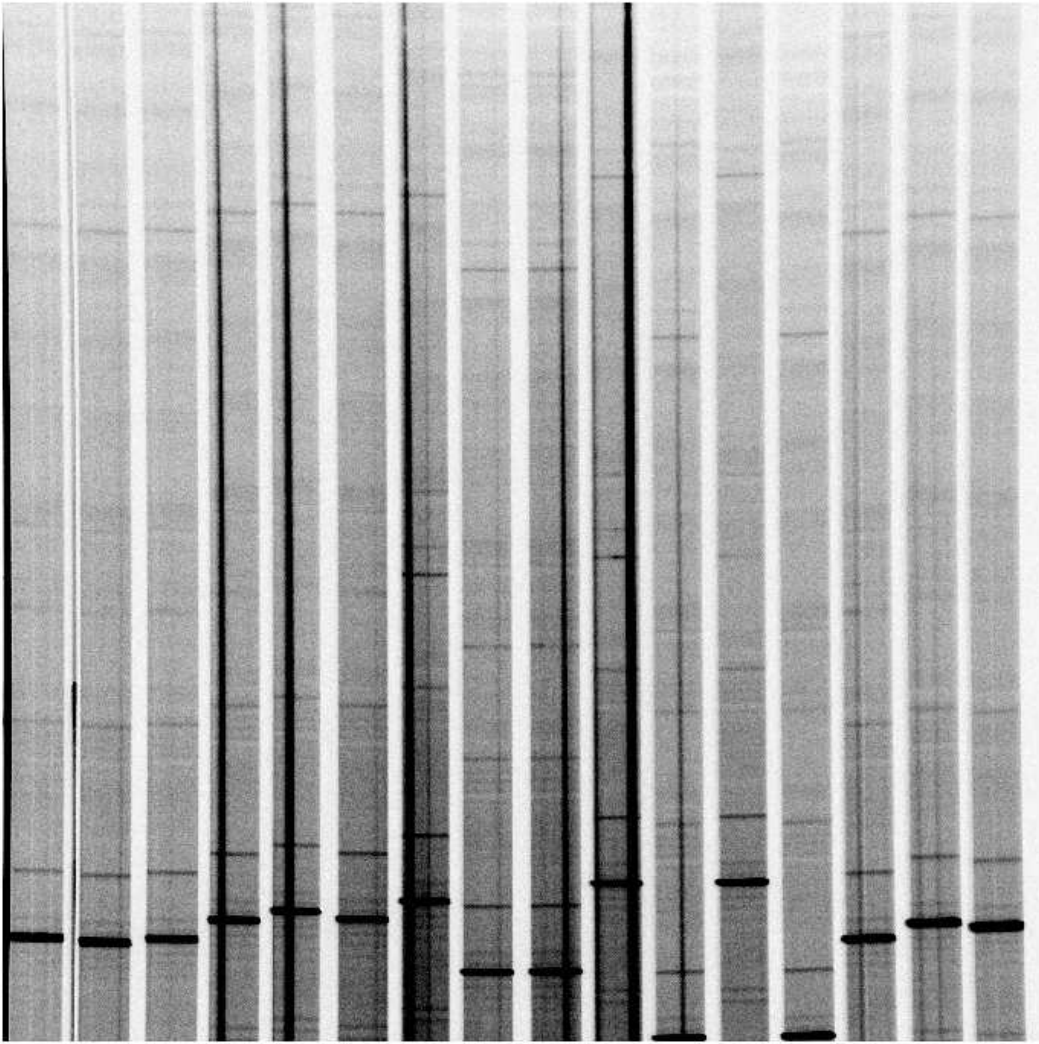}
\caption{Initial, non-reduced image obtained with SCORPIO in multi-slit mode.}
\end{figure*}

A log of observations of Lick standard strar is given in Table~1. Resulting spectra of ten
standard stars are shown in Fig.~2.

\begin{table}[hbt]
\caption{Observational log}
\begin{tabular}{lclc} \\ \hline
Object            & Date             & Exposure & Seeing         \\ \hline
HD132142 & 15.09.2004 &  10 c.     & 1"      \\
HD4744   & 15.09.2004 &  30 c.     & 1"       \\
HR1015   & 15.09.2004 &  10 c.     & 1"      \\
HD67767  & 16.12.2004 & 120 c.     & 3"      \\
HD72184  & 16.12.2004 & 20x2, 40 c. & 3"      \\
HD74377  & 15.12.2004 & 120, 240 c. & 3"      \\
HR0964   & 16.12.2004 & 20, 40 c.  & 3"      \\
HR3422   & 15.12.2004 & 120x2 c.   & 3"      \\
HR3427   & 15.12.2004 & 120x2 c.   & 3"      \\
HR3428   & 15.12.2004 & 120x2 c.   & 3"      \\ \hline
\end{tabular}
\end{table}

\begin{figure}[!t]
\psfig{figure=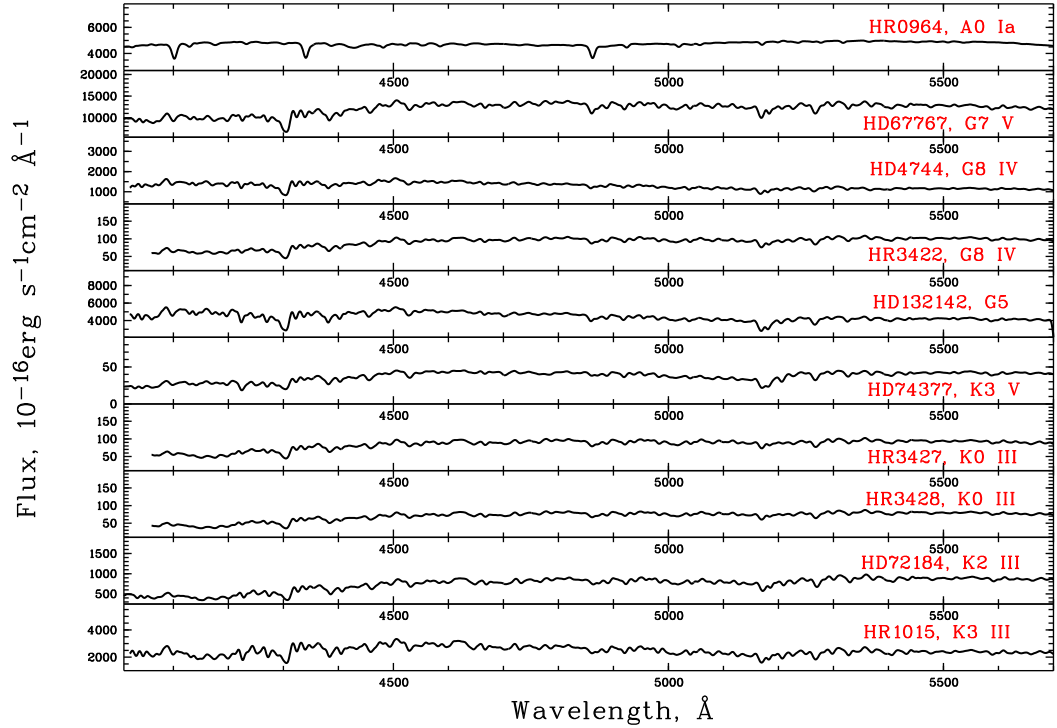,width=15cm,angle=-90}
\caption{Flux calibrated spectra of the Lick standard stars.}
\end{figure}

\begin{center}
{\bf \it Data reduction}
\end{center}

The data reduction was carried out using software packages written in the
Interactive Data Language (IDL). First, we performed the standard primary reduction of
our spectra: bias subtraction, cosmic particles removal, and
bad pixel replacement.

To properly determine the edges of our spectra for their
subsequent extraction, we remove the traces of bright light near slits 15 and 16 (see
Fig.~1) from the image. This cosmetic procedure was carried out on flat-field images. The
result of the subsequent determination of the edges of our 16 spectra was applied
to all spectra: the reference spectra and  the spectra of standard stars; and should be applied to
to all our globular cluster spectra.

Correcting the geometric field distortions on the reduced spectra in X and
Y directions (see Fig.~1) is the most important data reduction step.
We compute a polynomial of transformation between the final uniform
grid corresponding to the edges of spectra and the initial distorted one. The
grid is drawn with a uniform step along Y coordinate. The edges of the spectra serve as the reference points
for straightening the spectra.
To find these edges, we divide the original distorted image
into 10-pixel-long areas along the Y coordinate. These areas are summed in one-dimensional spectra
for which the intensity maxima are sought.
The derived coordinates of the maxima are
the edges of the spectra. The IDL POLY2D procedure is
used to find the polynomial transformation between two images represented by the corresponding reference points and to
calculate the succeeding transformation to the rectified image. The image
is corrected according to a polynomial of the form:
$$ x' = a(x,y)= \sum_{i=0}^{N} \sum_{j=0}^{N} P_{i,j} x^{j} y^{i}$$
$$ y' = b(x,y)= \sum_{i=0}^{N} \sum_{j=0}^{N} Q_{i,j} x^{j} y^{i},$$
where tensors P and Q contain polynomial coefficients. Each 2x2 tensor
contains $(N+1)^2$ elements. We apply this transformation formulae
to all the multi-slit spectra.

After the correction for the field distortion, we identify lines in the spectrum of a reference
He-Ne-Ar lamp, calculate dispersion curves, and perform  linearization in
wavelength space.

Since the instrumental conditions and the data reduction steps are identical for the flat fields
the spectra of the objects, we have the right to divide the object frames by the flat field
after the linearization of the spectra, not after bias subtraction and
cosmic particle hits removal, as is usually done. The multislit spectrum reduction procedure
is complex, because of the distortion
of spectra in the focal plane needs to be corrected.

\begin{figure*}[!th]
\vspace{-5cm}
\hspace{-1cm}
\psfig{figure=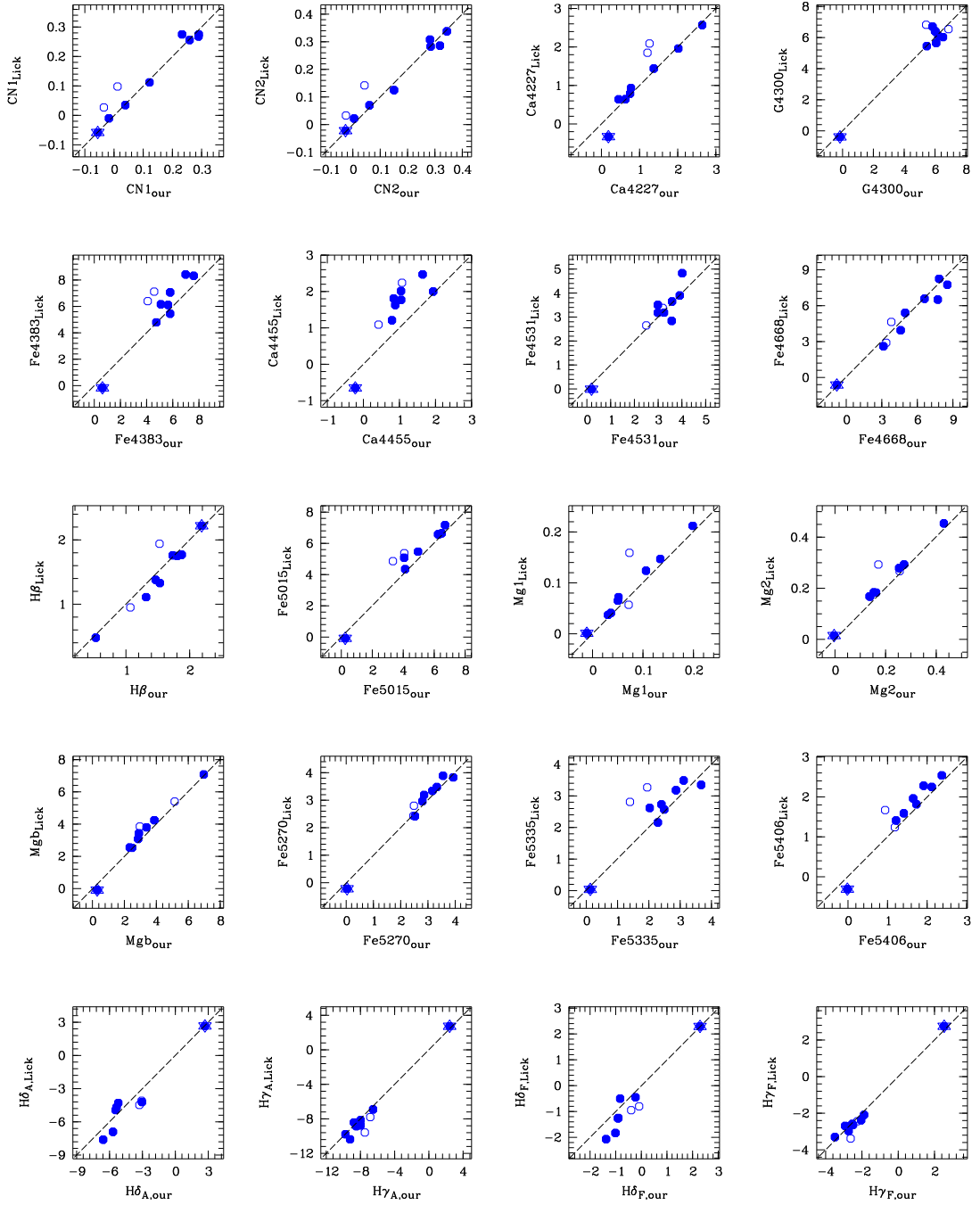,width=17cm}
\vspace{-4cm}
\caption{Comparison of passband measurements of our spectra and original Lick
data for 10 Lick standard stars. The dashed line shows the one-to-one relation.
Two outliers (HD132142 and HD4744) are shown as open circles and were 
excluded from calculations of the Lick transformation.
 The early A-type supergiant HR0964 (a star-like symbol) shows strongly
negative Ca4227, Ca4455 and Fe5406 index values in the calatog of Worthey et al. (1994),
and the lowest statistical weight was assigned to it.}
\end{figure*}
Subsequently, the sky spectrum is subtracted. Using a
the least-square method, the sky spectrum is interpolated in the spatial
direction and subtracted from the object spectrum. Unfortunately,
the sky subtraction method cannot be applied to bright Lick standard stars.
Even on short exposures the light of a bright star floods the entire slit.
However, in this case the sky background contributes only slightly to the final
spectrum. Occasionally, we detected scattered light between the slits. In this
case, it is more appropriate to subtract the sky spectrum taken from the other
(spatially close, but not neighbouring) slits. In the next section we show
that the Lick indices change only slightly (i.e. within the limits of the random measurement errors) after this
procedure. A similar approach to the sky
subtraction might be applied for the reduction of the integrated-light
spectra of extragalactic GCs, although the expected flux from the object in this case
will not flood the entire slit, giving rise to scattered light.

We use spectroscopic standard stars observed on the same night through the
medium slit to calibrate the fluxes of all Lick standard star spectra.
It should be noted that, by definition, the procedure of calculating
the Lick indices does not require the absolute flux calibration
(equations (1) and (2)). Hence, applying of the flux calibration to
standard star and the object spectra significantly reduces the non-linear
transformation offsets to the Lick system for indices with broad passband
definitions, since these indices are sensitive to the rate of change in the
continuum slope.

It is important to observe objects and standard stars in the same
instrumental conditions to correctly calculate the Lick indices. In
other words, the spectral sensitivity should be the same during our
observations. Therefore, we obtain the spectra for globular clusters and
standard stars using the same multi-slit mode.

\bigskip

\begin{center}
{\large \bf Transformation from the SCORPIO to the Lick system}
\end{center}

\bigskip

\begin{table}[!hbt]
\caption{Summary of the coefficients $\alpha$ and $\beta$ 
and the rms errors of the linear transformation to the standard system for all index measurements.
The fifth and sixth columns show units of the indices and mean bootstrapp
errors of the index values for 10 standard stars.
The last column shows a mean standard deviation of
index values for the stars with the indices measured on 13 spectra
obtained by subtracting the sky spectrum taken from different (not
neighbouring) slits.}
\begin{tabular}{lrrrcrcc} \\ \hline
Index       & $\alpha$   & $\beta$  & rms  & units & & $\sigma$(Index) & $\sigma$(Index)$_{msl}$  \\ \hline
CN1         & 0.036      & -0.0085  & 0.015& mag   & & 0.0022 &0.0027  \\
CN2         & 0.052      & -0.143   & 0.041& mag   & & 0.0040 &0.0018  \\
Ca4227      & 0.365      & -0.008    & 0.185&  \AA  & & 0.097  &0.0123  \\
G4300       & 0.786      & -0.086   & 0.713&  \AA  & & 0.105  &0.0198  \\
Fe4384      & -0.202     &  0.362   & 1.661&  \AA  & & 0.117  &0.0250  \\
Ca4455      & 0.650      & -0.001   & 0.324&  \AA  & & 0.120  &0.0085  \\
Fe4531      & -0.068     & 0.186    & 0.324&  \AA  & & 0.126  &0.0160  \\
Fe4668      & 0.223      & -0.054   & 0.731&  \AA  & & 0.141  &0.0199  \\
H$\beta$    &-0.339      & 0.124    & 0.053&  \AA  & & 0.141  &0.0070  \\
Fe5015      & 0.279      & 0.145    & 0.584&  \AA  & & 0.148  &0.0160  \\
Mg$_1$      & 0.016      & 0.014    & 0.034& mag   & & 0.0048 &0.0002  \\
Mg$_2$      & 0.018      &  0.007   & 0.011& mag   & & 0.0048 &0.0003  \\
Mgb         & 0.064      & 0.054    & 0.271&  \AA  & & 0.154  &0.0067  \\
Fe5270      & -0.287     & 0.198    & 0.186&  \AA  & & 0.157  &0.0088  \\
Fe5335      & 0.312      &  0.073   & 0.444&  \AA  & & 0.157  &0.0059  \\
Fe5406      & 0.239      & -0.018   & 0.150&  \AA  & & 0.158  &0.0051  \\
H$\delta_A$ & -0.647     & 0.035    & 1.340&  \AA  & & 0.178  &0.1600  \\
H$\gamma_A$ & -0.056     & 0.083    & 0.901&  \AA  & & 0.188  &0.0344  \\
H$\delta_F$ & -0.348     & -0.038   & 0.451&  \AA  & & 0.194  &0.0450  \\
H$\gamma_F$ &  -0.069    &  0.117   & 0.463&  \AA  & & 0.196  &0.0180  \\
\hline
\end{tabular}
\end{table}

Before measuring the indices, we degraded our spectra to the resolution of the
Lick system. The effective resolution (FWHM) of our spectra was
determined as a full width at half maximum (FWHM) of the corresponding autocorrelation function
divided by $ \sqrt 2$ (Tonry \& Davis, 1979). Our resolution correction
technique consisted of broadening our spectra to the Lick resolution with
a Gaussian filter with the dispersion $$ \sigma_{smooth}(\lambda) = \left(
\frac{{\rm FWHM} (\lambda)^2_{Lick}-{\rm FWHM}( \lambda)^2_{data}} {8 \ln 2} 
\right)^{1/2}, $$ 
taking into account the wavelength-dependent resolution of the
Lick system (see Worthey \& Ottaviani 1997).
Lick indices were measured on the spectra corrected for radial velocities.
Figure~3 shows the comparison between the Lick data and our index
measurements for all passbands. Least-square fits are used to parameterize
the deviations from the Lick system as a linear function of wavelength:
$$ EW_{cal} = \alpha + (1+ \beta) EW_{raw},$$ where $ EW_{cal}$ and
$EW_{raw}$ are calibrated and raw indices, respectively. Table 2 summarizes
the individual coefficients $\alpha$ and $\beta$. It is seen that our
instrumental system satisfactorily reproduces the standard Lick system, in
particular in the important indices such as Mg$b$, Mg$_{2}$, Fe5270,
Fe5335, and all Balmer indices.

 It is interesting to note, that $\sim$20\% of atomic metal line indices
have negative values for O, B, A spectral type stars in the
catalog of Worthey et al. (1994). This is probably due to the high
probability of chemical peculiarities for young hot stars
(see spectroscopic atlases: e.g. Chentsov et al. 2003, Albayrak et al. 2003).
Broad pseudocontinuum regions include many absorption line features
that are much more intense than the studied ones. Thus,
the flux per unit wavelength in the index passbands appear to be
lower than the corresponding fluxes in the
pseudocontinuum regions.

The influence of different sky subtraction methods on the Lick index measurements
is illustrated in the last column of Table~2. The indices were measured
for each standard star on 13 spectra obtained by subtracting the sky
spectrum taken from different (not adjacent) slits. It is seen from this table  that the
standard deviation does not exceed a few percent and is larger for the
Balmer lines, H$\delta_A$, H$\gamma_A$, H$\delta_F$, H$\gamma_F$, located near the edges of the spectra.

\begin{center}
{\bf \it Acknowledgements}
\end{center}
We thank S.N.~Dodonov for supervision of our observations, A.V.~Moiseev
for useful comments and V.P.~Mikhailov and for his help in providing the
observations, an anonymous referee and E.L.~Chentsov for useful comments. 
THP acknowledges support in form of an ESA Research Fellowship.

\bigskip
{\large \bf References}
\bigskip

{\noindent Albayrak B., Gulliver A. F., Adelman S. J., Aydin C., Kocer D.,
 A\&A, {\bf 400}, 1043 (2003)

\noindent Afanasiev V.L, Moiseev A.V., AstL, {\bf 31}, 216 (2005)

\noindent Brusual G., Charlot S., MNRAS, {\bf 344}, 1000 (2003)

\noindent Burstein,D., Faber S.M., Gaskell C.M., \& Krumm N.,
 ApJ, {\bf 287}, 586 (1984)

\noindent Chentsov E. L., Ermakov S. V., Klochkova V. G.,
Panchuk V. E., Bjorkman K. S., \& Miroshnichenko A. S., A\&A, {\bf 397}, 1035 (2003)

\noindent Geisler D., PASP, {\bf 100}, 687 (1988)

\noindent Faber S.M., ApJ, {\bf  179}, 731 (1973)

\noindent Puzia T.H., Saglia, R. P., Kissler-Patig, M. et al., A\&A, {bf 395}, 45 (2002)

\noindent Rich R.M., AJ, {\bf 95}, 828 (1988)

\noindent Sharina M.E., Sil'chenko O.K., Burenkov A.N., A\&A, {\bf 397}, 831 (2003)

\noindent Thomas D., Maraston C., Bender R., MNRAS, {\bf 339}, 897 (2003)

\noindent Tonry J., Davis M., AJ, {\bf 84}, 1511 (1979)

\noindent Rich R.M., AJ, {\bf 95}, 828 (1988)

\noindent Vazdekis A., ApJ, {\bf 513}, 224 (1999)

\noindent Worthey G.  ApJS, {\bf 95}, 107 (1994)

\noindent Worthey G., Ottaviani D.L., ApJS, {\bf 111}, 377 (1997)

\noindent Worthey G., Faber S.M., Gonzalez J.J., Burstein D., ApJS, {\bf 94}, 687 (1994)

\end{document}